\documentstyle[twocolumn,aps,eqsecnum]{revtex}
\begin{document}

\title
{Non-Fermi-Liquid Behavior of Compressible States of Electrons 
on the Lowest Landau Level}

\author
{\bf D.V. Khveshchenko}
\address
{NORDITA, Blegdamsvej 17, Copenhagen DK-2100, Denmark}
\maketitle

\begin{abstract}
\noindent
Experiments show that at even denominator fractions (EDF) 
($\nu=1/2, 3/4, 3/2$,...) the two-dimensional electron
gas (2DEG)
in a strong magnetic field becomes compressible, has no energy gap, and
demonstrates the presence of an ostensible Fermi surface (FS).
Since this phenomenon results from a minimization
of the interaction, rather than the kinetic energy,
the EDF states might well exhibit deviations from a
 conventional Fermi liquid (FL).    
We show that impurity scattering at EDFs
and its interference with electron-electron and
electron-phonon interactions 
provide examples of intrinsically non-Fermi-liquid (NFL) transport.
\end{abstract}
\vfill

A number of FL-like features 
exhibited by the strongly correlated EDF electronic states at $\nu\sim 1/\Phi$ ($\Phi=2,4,...$)  
\cite{exp} motivated the theoretical idea \cite{HLR} to describe these states 
as a new kind of Fermi liquid, which is formed by spinless fermionic quasiparticle named composite fermions
(CFs). On the mean field level the CFs, regarded as spin-polarized electrons
bound to $\Phi$ flux quanta, experience zero net field and occupy all states with momenta $k<k_{F,cf}=(4\pi n_e)^{1/2}$, where $n_e$ is the 2DEG density, below the effective CF FS. Formally, 
residual interactions of the CFs, as well as their interactions with charged impurities (remote
ionized donors sitting on a distance $\xi\sim 10^2 nm$ apart from the 2DEG) turn out to be essentially more singular than the original Coulomb ones. In the framework of the Chern-Simons theory of Ref.\cite{HLR} these interactions
appear as gauge forces, whose strength
 depends on the form of the electron interaction potential
$V_{ee}(q)$. In turn, the Coulomb potential of randomly distributed donors $V_{ei}(q)\sim 1/qe^{-q\xi}$ changes to that of random magnetic fluxes (RMF): $V_{ei}(q)\sim i({\bf p}\times {\bf q}/q^2)e^{-q\xi}$.

Conceivably, a 2D Fermi system governed by long-ranged 
gauge interactions of both dynamic (exchange by $\omega\neq 0$ gauge fluctuations) and static (RMF due to impurities and/or $\omega=0$ gauge fluctuations at $T>0$)
nature, could demonstrate quite unusual properties, and thereby provide an example of a genuine 2D NFL. 

Indeed, a naive attempt to proceed beyond mean field and
estimate perturbatively the effects of gauge fluctuations on the CF spectrum reveals
(even in a purely academic case of $T=0$ and no impurities)
a singular self energy, which behaves in the Coulomb case $V_{ee}(q)\sim 1/q$
 as $\Sigma_{cf}\sim \epsilon \ln\epsilon$, suggesting a divergent effective mass
$m^*_{cf}(\epsilon)\sim \ln\epsilon$. An even more singular $\Sigma_{cf}\sim \epsilon^{2/3}$ emerges
in case of a short-range interaction $V_{ee}\sim const$. Moreover, in a more realistic situation
with impurities present and/or $T>0$ the perturbative analysis fails to yield any finite self-energy
(formally, $\Sigma_{cf}=\infty$).

A non-perturbative eikonal calculation \cite{D1} shows a super-exponential decay of a CF wave packet
moving along its semiclassical trajectory: $<\Psi_{cf}(v_Ft,t)\Psi^{\dagger}_{cf}(0,0)>\sim \exp(-t\ln t)$
($\sim \exp(-t^2)$ in the short-range case), which implies a completely incoherent character
of a CF propagation, resulting in a failure of the naive "golden rule"-type estimates of $\Sigma_{cf}$
and other gauge-non-invariant quantities.

Therefore, an impressive
qualitative success of the mean field CF theory \cite{HLR} in explaining the experiments
\cite{exp}, where CFs seem to
propagate over distances of a few $\mu m$, requires a firm theoretical understanding.

It had been shown in \cite{KLW} that
the electrical current
relaxation processes, which correspond to smooth fluctuations of the ostensible CF FS,  can be safely 
described by means of the kinetic equation, where the singular self-energy and the Landau function-type terms
largely compensate each other. This implies that  
at small $q$ and $\omega$ the 
electromagnetic response functions of EDF states $K_{\mu\nu}(\omega, {\bf q})$
exhibit no singularities and can be computed
within the random phase approximation (RPA) \cite{SH}.

We arrived at the conclusions similar to those drawn in Ref.\cite{KLW} by applying to the EDF problem the geometrical method of the 2D bosonization developed in \cite{D2}.
In contrast to the earlier versions of the 2D bosonization the 
procedure of Ref.\cite{D2} facilitates an account of a finite FS curvature. The latter is crucially
important for a bosonic description of $D>1$ Fermi systems, and in particular, 
for calculating such physical quantities as Hall conductivity $\sigma_{xy}$
or diffusion thermopower
$S_d$, which simply vanish in the particle-hole symmetric limit of zero FS curvature.

The geometrical approach to the $D>1$ bosonization starts off with a choice of a particular basis of
coherent states $|\{ g({\bf p},{\bf q})\}>={\hat g}|0>$, which are generated by operators 
${\hat g}=\exp[i\sum_{{\bf p},{\bf q}}g({\bf p},{\bf q}){\hat n}({\bf p},{\bf q})]$ from the reference state
$|0>$ corresponding to the unperturbed circular FS of radius $k_{F}$,
in order to quantize the algebra of bi-linear spinless fermionic operators
${\hat n}({\bf p},{\bf q})=\Psi^{\dagger}_{\bf p}\Psi_{{\bf p}+{\bf q}}$
(2D analogue of the 1D algebraic structure known as $W_{\infty}$). 
The orbit of the group action  ${\hat g}|0><0|{\hat g}^{-1}$ can be parameterized in terms of the phase space "distribution function" $f({\bf r},{\bf p})$, which describes spatial fluctuations
of the occupied ("Luttinger") volume in momentum space. Some additional analysis is required to
identify all important kinds of FS fluctuations for a given Hamiltonian $H$. It turns out that in
the case of small angle scattering with transferred momenta being almost tangential to a FS
the most relevant fluctuations are those of the FS shape, and not of its profile. 
If only shape fluctuations are present, then $f({\bf r},{\bf p})$ amounts to
a support function $\Theta (k_{F}({\bf r})-p)$, and a local Fermi momentum 
${\bf k}_{F}({\bf r},\Omega)$ at the FS point $\Omega$ 
can be used as an unconstrained (obviously, bosonic) variable. The local density $\rho({\bf r})=\oint {d\Omega\over 4\pi} 
{\partial {\bf k}_{F}\over \partial\Omega}\times {\bf k}_{F}$ and current ${\bf J}({\bf r})
=e\oint {d\Omega\over 4\pi} 
{\bf v}_F{\partial {\bf k}_{F}\over \partial\Omega}\times {\bf k}_{F}$ (${\bf v}_F$ is given by the second functional
derivative of $H$: $\delta H=\epsilon_F
{1\over 2\pi}{\partial {\bf k}_{F}\over \partial\Omega}\times {\bf k}_{F}$,
$\delta\epsilon_F ={\bf v}_F\delta{\bf k}_F +\oint {d\Omega^{\prime}\over 4\pi} 
{\partial {\bf k}_{F}\over \partial\Omega^{\prime}}\times {\bf k}_{F}$) 
are, in general, non-linear functionals 
of ${\bf k}_F$. In the presence of an external field $\Delta B$ the equation of motion for
${\bf k}_{F}$ reads in the clean limit as
\begin{equation} 
{\partial {\bf k}_{F}\over \partial\Omega}\times (\partial_{t} -{\bf v}{\bf \nabla} +{\Delta B\over m^*}\partial_{\Omega}){\bf k}_{F}=0
\end{equation}
and serves as a counterpart of the kinetic equation, where a perturbation of the equilibrium distribution
function $f_0=\Theta (k_F-p)$ is proportional to a local FS displacement: $\delta f({\bf r}, {\bf p})\sim {\partial {\bf k}_{F}\over \partial\Omega}\times \delta{\bf k}_{F}$.

In the framework of the Chern-Simons theory of Ref.\cite{HLR} our method of the 2D 
bosonization enables one to obtain
 closed integral expressions for the irreducible CF polarization $\Pi_{\mu\nu}(\omega, {\bf q})$,
which is related to the correlator
$<k_F^{i}({\bf r}, \Omega, t)k_F^{j}({\bf 0}, \Omega^{\prime}, 0)>$. Thereby we confirm a number of results, which were
 obtained earlier in the "optical regime" $1/\tau_{tr}<<\omega<<E_F$ and $1/l<<q<<k_{F,cf}$ ($l=v_{F,cf}\tau_{tr}$ is the CF mean free path) while neglecting impurities \cite{SH,AIM}. For instance, 
the irreducible CF density polarization behaves similarly to that of free fermions with a finite mean field 
mass $m^*_{cf}\sim k_{F,cf}/e^2$, namely: $Re\Pi_{00} \sim m^*_{cf}$, $Im\Pi_{00} \sim m^*_{cf}\omega/k_{F,cf}q$.

The CF polarization 
still remains to be
related to $K_{\mu\nu}$, which requires an analogue
of the Silin's extension of the standard Landau FL theory onto the case of long-ranged interactions. 
In the problem at hand it yields the physical response functions in the RPA form \cite{HLR,SH}. Despite of the absence
of a small parameter, such as $e^2m^*_{cf}/k_{F,cf}$, the RPA form of the long wavelength
response is dictated by
the asymptotic Ward identities, which stem from an approximate particle number conservation at every
FS point $\Omega$.

In particular, the dynamic structure
factor has a form
$S(\omega,{\bf q})=Im K_{00}(\omega, {\bf q})\sim \omega q^3/(\omega^2+V^2_{ee}(q)q^6)$. The corresponding static structure factor $S_{EDF}(q)=\int d\omega S(\omega, {\bf q})\sim q^3\ln q$ has to be contrasted with the result $S_{FQHE}(q)\sim q^4$ for the incompressible FQHE states.

The response, however, becomes different from the FL-like at large $q$, and
the divergency of $m^*_{cf}$ does show up in processes, which correspond to rough
fluctuations of the CF FS \cite{KLW}. Such processes are responsible
for the SdH-type oscillations of $\rho_{xx}(B)=\rho_{xx}(0)+\sum_{k}\rho_k\cos (4\pi^2n_e/\Delta B)$ at 
$y=\tau_{tr}\Delta B/m^*_{cf}>1$. In the
semiclassical theory of magnetotransport $\rho_k$ can be related to a gauge field phase factor 
of a single CF making $k$ laps along its cyclotron orbit of radius
$R_{cf}=k_{F,cf}/\Delta B$: $\rho_k\sim <\exp( i\oint {\bf a}d{\bf r})>$. 
 
Accordingly, the experiments \cite{NFL} show that the corresponding Dingle plot for $\ln (\Delta\rho_{xx}(B)/\rho_{xx}(0))$ exhibits such NFL
features as a $\sim (\Delta B)^{-4}$ enveloping function at $T\rightarrow 0$ and 
a divergent $m^*_{cf}(\Delta B)\sim (\Delta B)^{-4}$ derived from the $T$-dependent
part in the regime $y>1$.

We also confirmed a strong enhancement of the density response at $q=2k_{F,cf}$,
which is another NFL feature. Our results, however, do not support 
the conjecture of a real divergency of the 
corresponding susceptibility in the particle-hole channel \cite{AIM}. If confirmed, it would make questionable the very stability of the CF metal against a CDW instability, which in this context can be identified 
with Wigner crystallization (WC). The experimental data do not show any signature of
the WC for EDFs at $\nu>1/4$ either. In addition, we found no pairing instability in the Cooper channel at total momentum zero. Therefore, at this stage one can argue that a pure system of CFs remains in a truly metallic state down to zero $T$.
 
Besides confirming the earlier observations, we extended our analysis onto realistic (impure) case, where the disorder effects are, by no means, negligible at small $\omega$ and $q$.
In fact, the remote donors, which supply electrons to the 2DEG and provide the dominating mechanism of scattering at low densities $n_e\sim 10^{11}cm^{-2}$, should be treated as an intrinsic element of the system.

Both the method of Ref.\cite{KLW} and our 2D bosonization provide a natural framework for 
account of disorder,
which can be introduced into Eq.(1) as a RMF: $\Delta B\rightarrow \Delta B + b({\bf r})$ with a white-noise correlator $<b({\bf q})b(-{\bf q})>=2\pi^2\Phi^2n_ee^{-2q\xi}$. 

First, we solve Eq.(1) while neglecting the residual gauge interactions among CFs \cite{DVK1}. At $\Phi>1$ the RMF problem
does not feature a small parameter, which would enable one to apply a customary
Born appoximation. However, under the condition
$k_{F,cf}\xi> \Phi$ one can resort on the eikonal-type solution for ${\bf k}_F({\bf r}, \Omega, t)$. In particular, this solution yields the CF conductivity
$
\sigma_{xx,cf}={e^2\over h} (k_{F,cf}\xi)\exp({\Phi^2/2})K_1(\Phi^2/2)
$
which appears to be more than twice as large as the result of the first Born approximation \cite{HLR}.  

In the presence of a uniform
field $\Delta B$ Eq.(1) yields a non-trivial
negative magnetoresistance (MR) $\Delta\rho_{xx}(\Delta B)/\rho_{xx}(0)\approx -0.06y^2$. This result implies that one should probably take into account the CF gauge interactions in order to reconcile the theory with experiment, which shows broad minima of $\rho_{xx}(B)$ in the vicinity of the primary EDFs.

At $T>0.1mK$ it suffices to include just the first order correction to the RMF solution,
which results from quantum interference between the RMF scattering and the CF gauge interactions \cite{DVK2}. 
In the framework of the kinetic equation it stems from corrections to the non-equilibrium CF density of states
and the non-local part of the collision integral. Higher order terms can only become relevant at very low $T$.

Proceeding this way, we obtain the correction to the tensor
of the CF conductivity tensor ${\hat \sigma}_{cf}$, which is related to the physical one as
${\hat \sigma}^{-1}_{cf}={\hat \sigma}^{-1}-(h/e^2\nu)\pmatrix{0 & 1\cr
-1 & 0}$, at finite
$y<1$ (before the onset of the SdH oscillations):
\begin{equation}
\Delta \sigma_{xx,cf}=(1-y^2)\Delta \sigma_{cf}, ~~~\Delta \sigma_{xy,cf}=2y\Delta \sigma_{cf} 
\end{equation}
where $\Delta\sigma_{cf}={2e^2\over h}\ln (T\tau_{tr})\ln (k_{F,cf}l)$ for a short-range $V_{ee}$, and
$\Delta\sigma_{cf}={2e^2\over h}\ln (T\tau_{tr})[\ln (k_{F,cf}l)+{1\over 4}\ln (T\tau_{tr})]$ in the Coulomb case.  These estimates are in a qualitative agreement with the experimental data from \cite{log}.

Compared to its zero field counterpart for ordinary electrons, the correction (2)
is enhanced and non-universal (it is larger in samples of higher mobility). It also yields a positive contribution to the MR, which is greater than the negative RMF term. Therefore, preceding the SdH regime, the overall MR in the vicinity of primary EDFs increases as $\sim (\Delta B)^2$. 

Also, the quantum interference corrections manifest themselves in the response of
EDF state to an applied thermal gradient.
In contrast to the case of zero field, the tensor of thermoelectric coefficients $\hat \eta$  
receives a $\ln T$ contribution, and
so does the low-$T$ diffusion thermopower (TEP) ${\hat S}_d={\hat \sigma}^{-1}{\hat \eta}$ 
at EDFs: $\Delta S_d \sim T\ln T$ \cite{DVK3}. Observation of
this effect should be, in principle, possible with the existing experimental techniques.

Thus, we obtained that when it comes to a coupling to another subsystem (impurities), 
the CF metal exhibits NFL properties even
in the hydrodynamic regime, which is 
described by the kinetic equation.

Another example of this sort is provided by the electron-phonon interaction at EDF.
In $GaAs$ at $T< 3-4K$ the only important is coupling  
via the piesoelectric (PE) potential. Therefore, one can treat phonons as bulk acoustic modes coupled to the local 2DEG density via the vertex
$M_{PE}({Q})\sim Q^{-1/2}$, where ${\bf Q}=({\bf q}, q_z)$ is a 3D phonon momentum. 
 
Summing up the RPA sequence of CF polarization diagrams we obtain that a PE potential 
generated by a lattice  displacement
induces both scalar and vector components of the gauge field, and therefore
 the CF-phonon vertex acquires both the density- and the current-like parts
\begin{equation}
M_{cf}=
{M_{PE}(Q)\over \epsilon_{cf}(\omega, q)}(1+ (2i\pi\Phi)F(q)
{{\bf v}\times{\bf q}\over q^2}\Pi_{00}(\omega,q)) 
\end{equation}
where the CF dielectric function $\epsilon_{cf}=1+F\Pi_{00}V_{e-e}+F^2(2\pi\Phi/q)^2\Pi_{00}\Pi_{\perp}$
depends on the formfactor $F(q)$ given in terms of a profile of the 2D quantum well.

Similar to the case of ordinary electrons the scalar part of (3)
undergoes dynamic screening, and therefore its effects appear to be
subleading compared to the vector part.   

 In the Bloch-Gruneisen regime $T<T_D=2uk_{F,cf}\sim 5K$ (here $u$ is a sound velocity) the phonon-limited mobility behaves as $\mu^{-1}_{cf-ph}\sim T^3$,
which has to be contrasted to the case of ordinary electrons $\mu^{-1}_{e-ph}\sim T^5$. This new behavior
was experimentally found in \cite{K}.

The above dependence, however, can only hold at $T_1=u/l<T<T_D$ where the
dynamics of CFs remains non-diffusive, whereas at lower $T$ 
effects of quantum interference between the impurity scattering and the electron-phonon interaction break down the Matthiessen's rule.

This low-$T$ regime, which is hardly accessible in the zero field case ($T_{1,el}\sim 1mK$), 
becomes essentially more relevant in the case of CFs, since the absolute value of the resistivity
at primary EDF is more than two orders of magnitude higher than 
at zero field, and therefore
the CF transport time $\tau_{tr}$ is much  shorter than the electronic one.

In the diffusive regime $ql<1$ corresponding to $T<T_{1,cf}\sim 300mK$ the processes of small momenta transfers  contribute
to $\mu^{-1}_{cf-ph}(T)$ as
 $\ln (T_{1,cf}/T)$ provided the ratio $u/v_{F,cf}$ is fairly small.
At $T<T_{2,cf}=u^2/v_{F,cf}l$ this correction ceases to grow logarithmically and shows only a $\sim T^2$ downward deviation from its $T=0$ value 
$\mu^{-1}_{cf-ph}\sim (h^2_{14}\epsilon_0^5/\rho u k_{F,cf}e^9\tau^3_{tr})\ln (T_{1,cf}/T_{2,cf})$,
where $h_{14}$ is a non-zero component of the PE tensor.

Our formalism also allows one to investigate other, more complicated, aspects of the electron-phonon
interaction at EDFs \cite{DVK4}. 

It was an anomalous increase of the surface acoustic wave (SAW)
attenuation at $\nu=1/2$ which provided the first experimental evidence of the compressible EDF states
\cite{exp} and inspired the formulation of the CF theory of Ref.\cite{HLR}. 

The SAW phonons are coupled to the 2DEG by virtue of the vertex
$|M_{SAW}(q)|^2\sim \int d{q_z}|M_{PE}(Q)|^2$, which remains finite at $q\rightarrow 0$.

It became
customary to express the SAW attenuation in terms of the momentum-dependent conductivity
$\sigma(q)=i\sigma_{M}(1-\epsilon(\Omega_q,q))$, where $\sigma_{M}={\epsilon_0u/2\pi}\sim
5\odot 10^{-7}\Omega^{-1}$:
$
\Gamma_q\sim |M^{SAW}(q)|^2 Im K_{00}(\Omega_q,q)\sim qIm{({1+i\sigma(q)/\sigma_M})^{-1}}
$.
At $ql>1$ one has  
$\sigma_{0}(q)\sim 1/q$ for ordinary electrons, 
 whereas the momentum-dependent conductivity  at EDF $\nu\sim 1/\Phi$ 
is inversely proportional to the CF quasiparticle conductivity: $\sigma_{\nu}(q)\approx (e^2\nu/h)^2/\sigma_{cf}(q)\sim q$. 
At small $q$ both physical conductivities  approach their static values, which typically
satisfy the relations:
$\sigma_{0}>>e^2/h>>\sigma_{\nu}\sim \sigma_{M}$. In this regime the SAW attenuation becomes linear in momentum:
$\Gamma_q=\gamma q$, and at EDF $\nu$ the coefficient $\gamma_{\nu}$  appears to be strongly
enhanced compared to its zero field counterpart $\gamma_{0}$:
\begin{equation}
\gamma_{\nu}/\gamma_{0}={\sigma_{\nu}\over {1+\sigma^2_{\nu}/\sigma^2_{M}}}{1+\sigma^2_{0}/\sigma^2_{M}\over \sigma_0}
\approx {\sigma_0\sigma_{\nu}\over {\sigma^2_M+\sigma^2_{\nu}}}>>1 
\end{equation}
We predict a similar enhancement 
to occur in other thermoelectric measurements, which 
probe the low-$T$ dynamics of CFs and their interactions with phonons. 
For instance, at $T>100mK$ the measured TEP
shows a non-linear dependence due to a phonon-drag contribution
$S_g$ resulting from the momentum transfer from phonons, which acquire a net flux of momentum in the presence
of a thermal gradient, to the CFs through their interaction. 

At zero field calculations based on the standard kinetic equation yield the 
result $S_{g,el}\sim T^4$, which holds in the clean regime
$T>T_1$. Below $T_1$ it crosses over to  $S_{g,el}\sim  T^3$. 
This can be viewed as the change from the momentum-dependent conductivity $\sigma_0(q)\sim 1/q$ to a
constant one.

On the contrary, in the case of CFs the kinetic equation gives 
$S_{g,cf}\sim T^2$ at $T>T_{1,cf}$  
and $S_{g,cf}\sim T^3$ at $T<T_{1,cf}$.
Remarkably, the exponent is higher in the dirty limit, as opposed to the
situation at zero field. This is a direct consequence
of the fact that in the clean regime the momentum-dependent
EDF conductivity $\sigma_{\nu}(q)$ grows linearly with momentum.

In the regime of strong disorder both the zero field and the EDF phonon-drag TEP
share the same temperature dependence $\sim T^3$. However, the prefactors are drastically different, and
their ratio $S_{g,cf}/S_{g,el}$ is equal to that
of the SAW attenuation coefficients (4).
 
The experimental data from \cite{T} show, according to the authors, an "only marginally weaker $T$-dependence for
CFs than for electrons", which suggests that the CFs are well in the disordered regime. The data, fitted in \cite{T} with
$S_{g,el}\sim T^{4\pm 0.5}$ and $S_{g,cf}\sim T^{3.5\pm 0.5}$, demonstrate a two order
of magnitude enhancement of the prefactor in the CF case.

It also follows from our analysis that the observed 
similarity of the $T$-dependence
of the phonon-drag TEP at zero field and at primary EDF \cite{T} 
is not inconsistent with the drastic difference in the corresponding phonon-limited mobilities \cite{K}, provided both systems are well in the disordered regime.

Another informative experimental probe of the electron-phonon interaction is provided by
measurements of an effective temperature of the 2DEG as a function of applied current
$T(I)\sim I^{2/\alpha}$, where the exponent $\alpha$ characterizes the  
energy loss rate due to phonon emission: $P\sim T^{\alpha}$. 

At zero field the kinetic equation gives in the clean limit
the standard result $P_{el}\sim T^5$, which is consistent with the inelastic electron-phonon scattering rate $\tau^{-1}_{in}\sim T^3$. In the disordered regime it changes to a lower power $P_{el}\sim T^4$. 

On the contrary, the situation at EDF again appears to be reversed: 
the power-law dependence
$P_{cf}\sim T^3$, which refers to the clean regime $T_{1,cf}<T<T_{D,cf}$ and implies the inelastic scattering rate $\tau_{in,cf}^{-1}\sim T$, changes to a greater power $P_{cf}\sim T^4$ in the dirty limit $T<T_{1,cf}$. 
Comparing the above results obtained in the regime of strong disorder we conclude that, just 
like the case of the phonon-drag TEP, the ratio $P_{cf}/P_{el}$ is given by the factor (4).

Recently the dependence $P\sim T^4$ was observed and discussed in the context of the transitions between adjacent quantum Hall effect plateaus (both integer and fractional) without any reference to CFs \cite{C}. 
It is tempting to identify the nearly two order of magnitude enhancement ( compared to the zero field case)
of the emission
rate at the transition between $\nu=1/3$ and $\nu=2/5$, with the 
CF behavior governed by the EDF state at $\nu=3/8$.  
 A systematic experimental verification of this conjecture could lend an additional support for the CF theory.
  
To summarize, we study the NFL properties of realistic (impure) compressible EDF states
by using the kinetic equation written in terms of local deformations of the CF FS.
If temperatures are not very low ($T>0.1mK$), then, at first hand, one has to take into account the effects of Coulomb impurities, which play a role
of the long-range correlated RMF in the CF representation. To this end, we
find the explicit semiclassical
solution of the RMF problem beyond the first Born approximation. At small deviations from the primary EDFs
this solution yields the quadratic negative MR.
The effects of the residual CF interactions manifest themselves as quantum
interference corrections. These give rise to the enhanced non-universal $\ln T$ terms in such transport quantities as the electrical
resistivity or the diffusion TEP, and appear to be responsible for the observed broad minima of the MR
at the primary EDFs. We also carried out a comparative analysis of the effects of electron-phonon interactions 
at zero field and at EDFs.
We show that in the latter case the acoustic phonon contribution to the electronic 
mobility, the SAW attenuation, the phonon-drag TEP, and the energy loss rate for hot electrons, depending on temperature, either all contain smaller powers of $T$, or are enhanced by the same numerical factor (4).

\end{document}